\documentclass[twocolumn,floatfix,showpacs,preprintnumbers,amsmath,amssymb,prl]{revtex4}

\usepackage{graphicx}
\usepackage{dcolumn}
\usepackage{bm}

\begin{document}

\input{epsf}

\title{Field-Induced Ising Criticality and Incommensurability 
in Anisotropic Spin-1 Chains} 

\author{Y.-J.\ Wang}

\affiliation{Max-Planck-Institut f\"ur Physik Komplexer Systeme, 
N\"othnitzer Stra\ss e 38, 01187 Dresden, Germany}

\date{\today}

\begin{abstract}

Quantum phase transitions induced by an external magnetic field in the 
Haldane-gapped spin-1 chains are studied in a fermionic field-theoretic 
description of the model. In the case with broken axial symmetry, two 
transitions occurs consecutively as increasing the field: 
$h_c=(m_1 m_2)^{1/2}$ (with $m_1$ and $m_2$ the transverse masses) 
is the critical Ising point and $\tilde{h}_c^*\, (> \!h_c)$ is the 
commensurate-incommensurate transition point. Although the latter is 
thermodynamically indiscernible, its property is revealed in the spin 
correlation functions (both uniform and staggered) which are calculated 
upon expressing the correlations of coupled Ising pair fields in terms 
of block-Toeplitz determinants. 

\end{abstract}


\maketitle 

It is now well established that a one-dimensional (1D) quantum 
antiferromagnet with spin-1 possesses a (Haldane) gap in the excitation 
spectrum which separates the first excited triplet states from the singlet 
ground state \cite{hald83}. 
The elementary excitations are coherent antiferromagnetic massive magnons 
appearing around $q\!=\!\pi$ and carrying quantum number $S\!=\!1$. If an 
external magnetic field is introduced into the Haldane system, the transverse 
modes couple to the field and phase transition may occur when the field 
overcomes the gap \cite{aff90,tsv90,sak91,sach94,fur99}. Experimentally, 
there have indeed been observed phase transitions in some quasi-1D compounds 
by measuring thermodynamic quantities such as magnetization 
\cite{ren87,aji89,hon01} 
and specific heat \cite{hon98,hon01}, as well as neutron scattering peak 
\cite{ren87,chi91,chen01}. 

For an isotropic chain or the field applied having 
axial symmetry, a commensurate-incommensurate (C-IC) transition \cite{dzh78} 
takes place when the field surpasses the transverse mass, and the system 
enters the massless Luttinger liquid phase. Physically, this is understood 
in either bosonic \cite{aff90,fur99} or fermionic language \cite{tsv90}. 
In the former description, the transition is associated with 
the condensation of the bosons; while in the latter, the fermions begin to 
fill the conduction band when the field (as chemical potential) exceeds 
the gap. Aside from magnetization, the external field induces a planar 
quasi-long-range order in the staggered spin correlations. The correlation 
functions in the Luttinger liquid phase have also been calculated 
\cite{fur99}. However, in most real materials the spin 
rotational symmetry is hardly preserved in presence of single-ion 
anisotropies, and thus the triplet masses are at least slightly different. 
Considering the Luttinger liquid requires a U(1) symmetry and cannot survive 
a broken axial symmetry, in a general asymmetric case, the phase transition 
reduces to be of Ising-type \cite{aff90}. 
In this Letter, we report the results for the Ising scenario, in particular, 
the spin correlation functions in all phases. 

We take a field-theoretic model due to Tsvelik \cite{tsv90} for 
the anisotropic Haldane chain: 
$
{\cal H}_{\text{S}}(x) \!=\! \sum_{\alpha=1,2,3} [ -\frac{iv}{2} \big( 
\xi_R^\alpha \partial_x \xi_R^\alpha - \xi_L^\alpha \partial_x \xi_L^\alpha 
\big) - im_\alpha \xi_R^\alpha \xi_L^\alpha ] 
$, 
where $v$ is the characteristic velocity of magnons, and the triplet 
$\mbox{\boldmath $\xi$}_\nu \!=\! (\xi_\nu^1, \xi_\nu^2, \xi_\nu^3)$ 
are chiral ($\nu\!=\!R,L$) Majorana (real) fermion fields. The masses, 
$m_\alpha\, (>\!0)$, are considered as phenomenological parameters accounting 
for the anisotropy. Although the neglected marginal current-current 
interactions are essential to give rise to field dependent exponents in the 
Luttinger liquid phase, their role is exhausted in the Ising transition case by 
renormalizing the masses and velocity. Since the system is now massive 
everywhere except right at the critical point, the effect of interactions is 
at most shift the transition point without changing the universal property 
significantly. 
In this representation, the spin density, as a sum of uniform 
and staggered parts, is expressed as 
$
\mbox{\boldmath $S$} \!=\! \mbox{\boldmath $I$} + (-1)^n \gamma 
\mbox{\boldmath $N$}  
$
($\gamma$ being a constant). The uniform part 
$\mbox{\boldmath $I$} 
\!=\! \mbox{\boldmath $I$}_R + \mbox{\boldmath $I$}_L$ with the chiral 
currents 
$
\mbox{\boldmath $I$}_\nu \!=\! -\frac{i}{2} \mbox{\boldmath $\xi$}_\nu 
\mbox{\boldmath $\times$} \mbox{\boldmath $\xi$}_\nu 
$
satisfying the SU(2)$_2$ Kac-Moody algebra and having the scaling dimensions 
$(0,1)/(1,0)$ for $\nu\!=\! R/L$. The staggered part is associated with the 
primary field with scaling dimension $(\frac{3}{16}, \frac{3}{16})$: 
$
\mbox{\boldmath $N$} \!=\! (\sigma_1 \mu_2 \mu_3, \mu_1 \sigma_2 \mu_3,
\mu_1 \mu_2 \sigma_3) 
$, 
where $\sigma_\alpha$ and $\mu_\alpha$ are, respectively, order and disorder 
Ising variables. 

The magnetic field coupling to the uniform spin density contributes a Zeeman 
energy as 
$
{\cal H}_{\text{M}}(x) \!=\! h I^z(x) \!=\! -ih \big(\xi_R^1 \xi_R^2 
+ \xi_L^1 \xi_L^2 \big) 
$, 
which apparently only mixes first two (transverse) species of Majoranas, 
and the third one drops out of the system as a free massive Majorana, or 
equivalently, off-critical Ising model. Therefore, we need only concentrate 
on the Hamiltonian 
\begin{eqnarray} 
{\cal H}(x) &\!=\!& \sum_{\alpha=1,2} \Big[ -\frac{iv}{2} \big( 
\xi_R^\alpha \partial_x \xi_R^\alpha - \xi_L^\alpha \partial_x \xi_L^\alpha 
\big) -im_\alpha \xi_R^\alpha \xi_L^\alpha \Big] \nonumber \\ 
&& \quad\quad -ih \big(\xi_R^1 \xi_R^2 + \xi_L^1 \xi_L^2 \big) \,. \label{ham_c12} 
\end{eqnarray} 
Denoting $m\!=\! \frac{1}{2}(m_1 + m_2)$ and 
$\Delta\!=\! \frac{1}{2}(m_1 - m_2)$ 
and combining two Majoranas into a Dirac fermion, model (\ref{ham_c12}) 
actually describes a charge density wave-Cooper pairing-chemical potential 
($m$-$\Delta$-$h$) system in 1D. When bosonized, it becomes a sine-Gordon model 
with extra terms of dual field and field derivative. 

While the uniform spin correlations are easy to compute, it is not the case 
for the staggered part, since they are composed of Ising order and disorder 
fields, which are related to the Majorana fermions in a highly non-trivial way 
(Jordan-Wigner transformation). To exploit the full advantage of the 
solvableness of the model, we regularize (\ref{ham_c12}) on a 1D 
lattice, on which the relations between the Ising variables and the fermions 
can be unambiguously established. 
By using the relations between the Majorana fields and the lattice Dirac fermion:
\begin{equation} 
\Big\{ \begin{array}{l} \xi_R \to -\frac{1}{\sqrt{2a}} (e^{i\pi/4} 
c_n^\dagger + e^{-i\pi/4} c_n) \,, \\ \xi_L \to +\frac{1}{\sqrt{2a}} (e^{-i\pi/4} 
c_n^\dagger + e^{i\pi/4} c_n) \,, \end{array} \label{dcmp} 
\end{equation}
we have the following lattice version of two coupled quantum Ising chains 
(QICs) (or $m$-$\Delta$-$h$ model): 
\begin{equation} 
H = H_{\text{QIC}}^{[a]} +  H_{\text{QIC}}^{[b]} - ih \sum_n 
(a_n^\dagger b_n - b_n^\dagger a_n) \,, \label{CQICs} 
\end{equation}
where 
$
H_{\text{QIC}}^{[c]} \!=\! \frac{J}{2} \sum_{n} [ 
(c_n^\dagger - c_n)(c_{n+1}^\dagger + c_{n+1}) 
- \lambda_c (c_n^\dagger -c_n) (c_n^\dagger + c_n) ] 
$ with parameters $J\!=\! \frac{v}{a}$ and $\lambda_c \!=\! 1 + \frac{m_c}{J}$ 
($c \!=\!(1,2) \!=\!(a,b)$). 
Without loss of generality, we assume $m_1 \!\ge\! m_2 \!>\!0$ (so that 
$m \!>\! \Delta \!\ge\! 0$ and $\lambda \!>\!1$) and $h\!\ge\! 0$ unless 
otherwise stated. The units are so chosen that $J\!=\!a\!=\!1$. 

Model (\ref{CQICs}) is diagonalized through a unitary transformation and the 
spectrum is found to be 
\begin{equation} 
E_k^\pm = \big(\Theta_k^2 \pm 2 \Gamma_k^2 \big)^{1/2} \,,  
\end{equation} 
where $\Theta_k^2 \!=\! \Lambda_k^2 + \Delta^2 + h^2$ and 
$\Gamma_k^2 \!=\! \big(\Delta^2 \epsilon_k^2 + h^2 \Lambda_k^2 \big)^{1/2}$ 
with $\Lambda_k^2 \!=\! \epsilon_k^2 + \sin^2k$ and 
$\epsilon_k \!=\! \cos k -\lambda$. There exist totally {\em six} (real) 
correlation functions for the fermions: 
\begin{eqnarray} 
C_1^c(n) = \langle c_n^\dagger c_0 \rangle \,, \quad && 
C_2^c(n) = \langle c_n^\dagger c_0^\dagger \rangle \,, \quad (c=a,b) 
\nonumber \\  
C_3(n) = i \langle a_n^\dagger b_0 \rangle \,, \quad && 
C_4(n) = -i \langle a_n b_0 \rangle \,. \label{fmncf} 
\end{eqnarray} 
The explicit forms of these correlations are obtained in a variational 
approach, and all of them are 1D integrations over momentum. To calculate 
the correlations of the coupled Ising fields, we need further 
\begin{eqnarray} 
G_n^{a,b} &\!=\!& -\delta_{n0} + 2 [C_1^{a,b}(n) - C_2^{a,b}(n)] \,, \quad 
\text{(real)} \nonumber \\ 
F_n &\!=\!& 2i [C_3(n) - C_4(n)] \,. \qquad\quad \text{(imaginary)} 
\label{GnFn} 
\end{eqnarray} 

As is well-known, the correlation functions of the QIC can be expressed as 
the so-called Toeplitz determinants. A Toeplitz matrix has its elements 
parallel to the diagonal the same value, i.e., it takes the form 
$\hat{T} \!=\! [f_T(i-j)]$. 
We find that for the two QICs coupled in the present fashion, the correlation 
of a pair of Ising fields can be cast into a block-Toeplitz form with a doubled 
size. To this purpose, we introduce following {\em seven} $r \!\times\! r$ 
Toeplitz matrices: 
\begin{eqnarray} 
\hat{G}_\sigma^{a,b}(r) = [-G_{i-j+1}^{a,b}] \,, &&  
\hat{G}_\mu^{a,b}(r) = [G_{i-j}^{a,b}] \,, \nonumber \\ 
\hat{F}_\sigma(r) = [-F_{i-j+1}] \,, &&  
\hat{F}_\mu(r) =  [F_{i-j}] \,, \nonumber \\ 
\hat{\bar{F}}_\sigma(r) = [-F_{-i+j-1}] \,,  && \label{TpDt} 
\end{eqnarray} 
where the indices $(i,j)$ run from $1$ to $r$, and the matrix elements 
$G_n^{a,b}$ and $F_n$ are given by Eqs.~(\ref{GnFn}). Then, the Ising 
pair-field correlations like ${\cal C}_{\sigma\mu}^{ab}(r) \!\equiv\! \langle 
\sigma_{1} \mu_{2}(n) \sigma_{1} \mu_{2}(n+r) \rangle$ reduce 
to $(2r) \!\times\! (2r)$ block-Toeplitz determinants: 
\begin{eqnarray} 
{\cal C}_{\sigma\mu}^{ab}(r) &\!=\!& \det\bigg[ \begin{array}{cc} 
\hat{G}_\sigma^a(r) & -\hat{F}_\sigma(r) \\ \hat{F}_\mu(r) & 
[\hat{G}_\mu^b(r)]^T \end{array} \bigg] \,, \nonumber \\ 
{\cal C}_{\mu\sigma}^{ab}(r) &\!=\!& \det\bigg[ \begin{array}{cc} 
\hat{G}_\sigma^b(r) & -\hat{\bar{F}}_\sigma(r) \\ {[}\hat{F}_\mu(r){]}^T & 
[\hat{G}_\mu^a(r)]^T \end{array} \bigg] \,, \nonumber \\ 
{\cal C}_{\mu\mu}^{ab}(r) &\!=\!& \det\bigg[ \begin{array}{cc} 
\hat{G}_\mu^a(r) & -\hat{F}_\mu(r) \\ \hat{F}_\mu(r) & 
[\hat{G}_\mu^b(r)]^T \end{array} \bigg] \,, \nonumber \\ 
{\cal C}_{\sigma\sigma}^{ab}(r) &\!=\!& \det\bigg[ \begin{array}{cc} 
\hat{G}_\sigma^a(r) & -\hat{F}_\mu(r) \\ \hat{F}_\mu(r) & 
[\hat{G}_\sigma^b(r)]^T \end{array} \bigg] \,. \label{CBT} 
\end{eqnarray} 

With all these provisions, the spin correlations are readily obtained. For the 
uniform part, ${\cal S}_I^\alpha(r) \!\equiv\! \langle I^\alpha(n) 
I^\alpha(n+r) \rangle$: 
\begin{eqnarray} 
{\cal S}_I^{x,y}(r) &\!=\!& \delta_{r0} \big[ C_1^{b,a}(0) + C_1^c(0) \big] \nonumber \\ 
&& -2 C_1^{b,a}(r) C_1^c(r) -2 C_2^{b,a}(r) C_2^c(r) \,, \nonumber \\ 
{\cal S}_I^z(r) &\!=\!& \delta_{r0} \big[ C_1^a(0) + C_1^b(0) \big] \nonumber \\ 
&& -2 C_1^a(r) C_1^b(r) -2 C_2^a(r) C_2^b(r) \nonumber \\ 
&& - 2 [C_3(r)]^2 + 4 [C_3(0)]^2 + 2 [C_4(r)]^2 \,. \label{SI} 
\end{eqnarray} 
Here $C_{1,2}^c(r)$ are for the third (decoupled, off-critical) QIC. 
The extra terms in ${\cal S}_I^z(r)$ is due to the admixture of the first two 
QICs in the presence of the magnetic field. Actually, $4[C_3(0)]^2$ is the 
magnetization squared. For the staggered part, ${\cal S}_N^\alpha(r) 
\!\equiv\! \langle N^\alpha(n) N^\alpha(n+r) \rangle$: 
\begin{eqnarray} 
{\cal S}_N^x(r) &\!=\!& {\cal C}_{\sigma\mu}^{ab}(r)\, {\cal C}_{\mu}^c(r) \,, 
\quad 
{\cal S}_N^y(r) = {\cal C}_{\mu\sigma}^{ab}(r)\, {\cal C}_{\mu}^c(r) \,, 
\nonumber \\ 
&& \qquad {\cal S}_N^z(r) = {\cal C}_{\mu\mu}^{ab}(r)\, {\cal C}_{\sigma}^c(r) 
\,, \label{SN} 
\end{eqnarray} 
where ${\cal C}_{\sigma, \mu}^c(r)$ are Ising correlations for the 
third decoupled QIC, which bear simple Toeplitz forms ($r\!\times\! r$ 
matrices). 

\begin{figure}[t]
\begin{picture}(0,136)
\leavevmode\centering\includegraphics{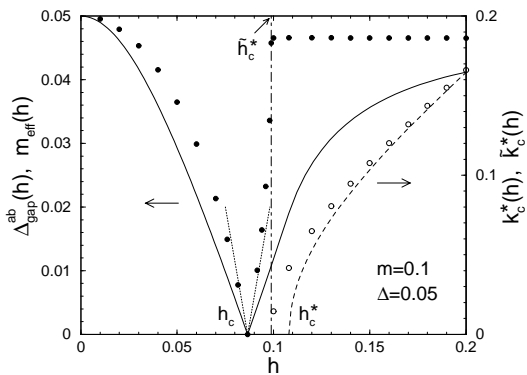}
\end{picture}
\caption{The energy gap ($\Delta_{\text{gap}}^{ab}$) of the coupled QICs 
($m$-$\Delta$-$h$ model) as a function of $h$ (the solid line). 
The gap vanishes at $h_c=0.0866025$ for $m=0.1$ and $\Delta=0.05$. 
Above $h_c^*$ (=0.108080), the dashed line is for the wavevector at which 
the gap locates. The (numerically extracted) solid circles represent the 
effective mass, $m_{\text{eff}}$, that governs all correlation functions; 
and the open circles are for the true incommensurate wavevector, 
$\tilde{k}_c^*$, that enters the correlation functions in the incommensurate 
phase. The vertical dash-dotted line at $\tilde{h}_c^*=0.098664$ is the true 
C-IC boundary. The wedge (dotted line) stuck to the $h_c$ point represents 
the mass deduced from the effective Ising theory.}
\label{fig:eng_gap} 
\end{figure}

The fundamental and instructive quantity is the energy gap. The gap of 
the first two coupled QICs ($m$-$\Delta$-$h$ model) is determined by 
$\Delta_{\text{gap}}^{ab} \!=\! \min(E_k^-)$. In axially symmetric case 
($\Delta \!=\! 0$), the gap decreases linearly: $\Delta_{\text{gap}}^{ab}(h) 
\!=\! (m-h) \theta(m-h)$. In general case, the gap function is shown in 
Fig.~\ref{fig:eng_gap} (solid line),  
which has the following features: 
When $h$ is small, it diminishes from $m_2$ (the smaller mass of the two) 
as $\Delta_{\text{gap}}^{ab}(h) \!\approx\! m_2 
- \frac{h^2}{2\Delta}$. In the vicinity of the critical point, 
$
h_c \!=\! (m^2-\Delta^2)^{1/2} \!=\!  (m_1 m_2)^{1/2} 
$, 
the gap vanishes linearly at both sides as 
$
\Delta_{\text{gap}}^{ab}(h) \!\approx\! \frac{h_c}{m} |h- h_c| 
$. 
When the magnetic field is further increased to $h_c^* \!\approx\! h_c 
+ \frac{\Delta^2}{\lambda m}$, 
the gap position in momentum space begins to shift from $k\!=\!0$ to 
$k_c^*(h) \!\approx\! \big[ (\frac{2m}{\lambda})^{1/2} 
+ \frac{(\lambda^2+2)\Delta^2}{(2m)^{3/2} \lambda^{5/2}} 
\big](h-h_c^*)^{1/2}$. 
As a result, a gapful incommensurate phase is expected at large $h$.
On the other hand, it is shown that when $h>h_c^*$ and for small $\Delta$, 
the gap $\Delta_{\text{gap}}^{ab} \!\sim\! \Delta$. 
Therefore, in the axially symmetric case, the gap closes in this region. 

The identification of the Ising criticality at $h_c$ follows from the linearly 
vanishing behavior of the gap and the diagonalized Hamiltonian. Since $E_k^+$ 
always remains massive, the modes that associated drop effectively out of the 
low-energy theory, which is resultingly an Ising theory keeping only half of 
the original degrees of freedom. By expanding the spectrum around the 
transition point and for long wavelengths, 
$
{E_k^-} \!\approx\! J_{\text{eff}} \big( k^2 + m_{\text{eff}}^2 \big)^{1/2} 
$, where $J_{\text{eff}} \!=\! \frac{\Delta}{m} <1$ is a new (smaller) energy 
scale for the effective theory with an effective mass $m_{\text{eff}} \!=\! 
\frac{h_c}{\Delta} |h-h_c|$ (the dotted wedge connected to the Ising point in 
Fig.~\ref{fig:eng_gap}). 

\begin{figure}[b]
\begin{picture}(0,106)
\leavevmode\centering\includegraphics{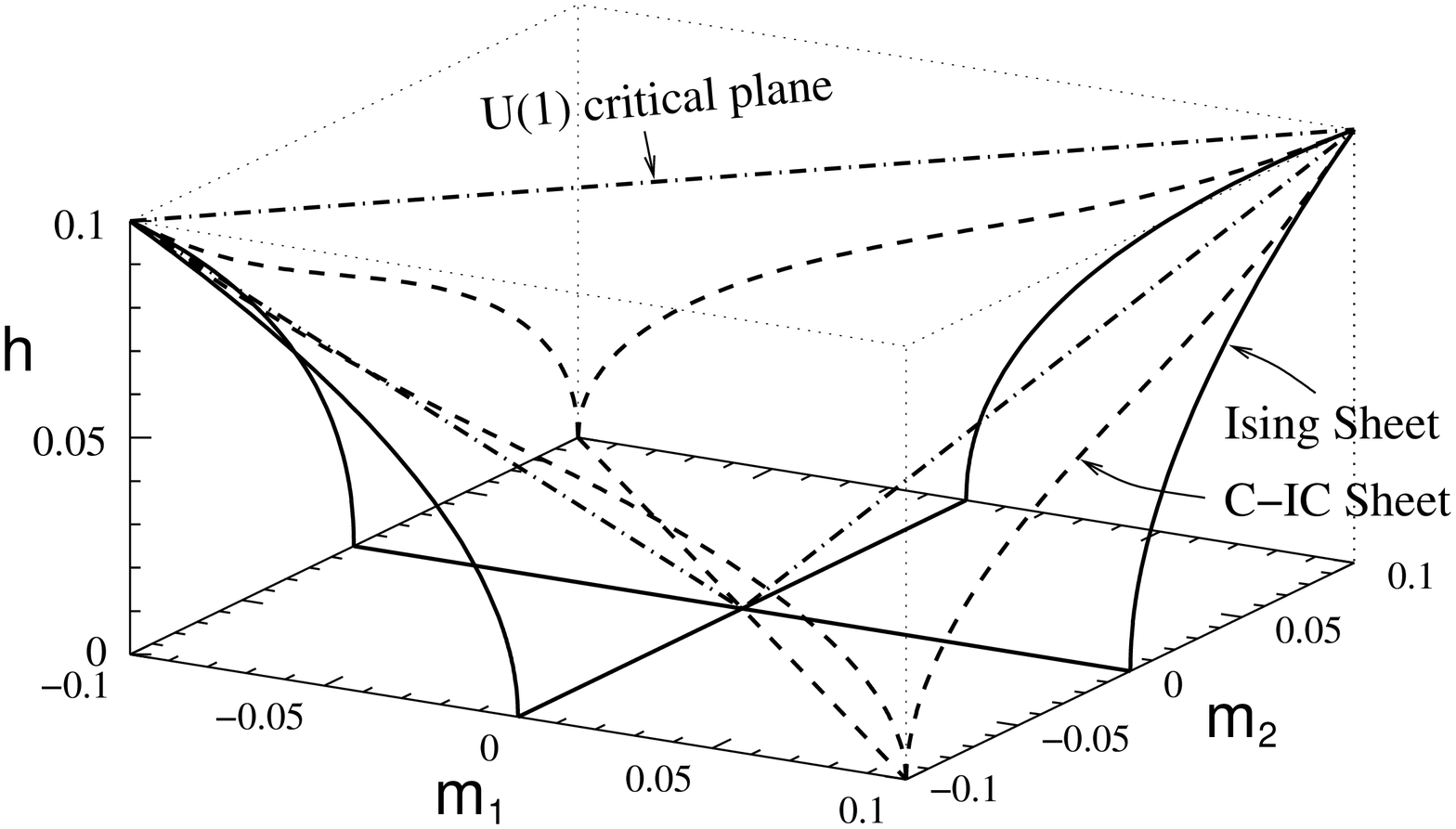}
\end{picture}
\caption{Phase diagram of the coupled QICs ($m$-$\Delta$-$h$ model). 
The sheets enclosed by the solid lines belong to the Ising criticality, and 
those by the dashed lines are the C-IC transition sheets. The triangle with 
dash-dotted sides stands for the incommensurate U(1) critical plane 
(Luttinger liquid).}
\label{fig:phase_dg} 
\end{figure}

The emergent $T\!=\!0$ phase diagram is shown in 
Fig.~\ref{fig:phase_dg},  
in which we release the constraint on the relative values and signs of the 
masses, so that it covers the whole parameter ranges for the $m$-$\Delta$-$h$ 
model (\ref{ham_c12}) or (\ref{CQICs}). The critical Ising sheets, determined 
by the condition $h=h_c(m_1, m_2)$, arise from the trivial Ising lines 
($m_1=0$ and $m_2=0$ at $h=0$), and merge the C-IC sheets, pinned down by 
$h=h_c^*(m_1, m_2)$, at the U(1) symmetric line $h=m_1=m_2$, i.e., the usual 
C-IC transition line in the axially symmetric case ($\Delta=0$). Above the  
transition only the system with the U(1) symmetry is critical.

Unfortunately, although $\Delta_{\text{gap}}^{ab}$ and $k_c^*$ are 
enlightening and analytically expressible, neither of them is the true 
quantity that controls the asymptotic behaviors of the correlation functions. 
We have to numerically extract the corresponding effective mass 
$m_{\text{eff}}$ (i.e., inverse of the correlation length) and effective 
incommensurate wavevector $\tilde{k}_c^*$ from the fermion correlation 
functions, Eqs.~(\ref{fmncf}), by evaluating the integrals. For comparison, 
we show these two quantities in the same Fig.~\ref{fig:eng_gap} (the solid 
and open circles). As a result, the real C-IC transition point $\tilde{h}_c^*$ 
locates between $h_c$ and $h_c^*$. We also see from the figure that the 
effective mass, which is in general greater than the gap of the system, 
deviates significantly from the gap above the Ising transition. The mass grows 
rapidly until the system enters the incommensurate phase where the mass 
becomes field independent. In the vicinity of the Ising transition, the 
linearized effective mass (dotted wedge) asymptotically coincides with the 
numerical mass values (solid circles). The dissociation between the gap and 
mass is in fact due to the external field which breaks the Lorentz invariance 
of the model.

\begin{figure}[t]
\begin{picture}(0,222)
\leavevmode\centering\includegraphics{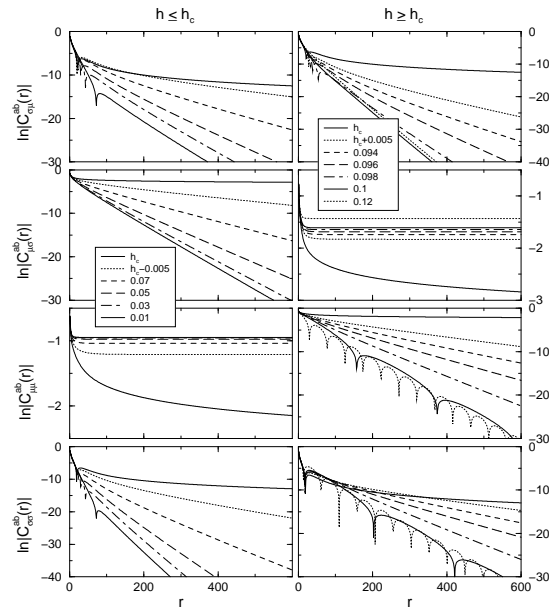}
\end{picture}
\caption{Four Ising pair correlation functions at various magnetic field 
strengths. The parameters are same as in Fig.~\ref{fig:eng_gap}.}
\label{fig:Iscf} 
\end{figure} 
By combining Eqs.~(\ref{CBT}), (\ref{TpDt}), and (\ref{GnFn}), and the 
known values of functions $C_1^a(r)$--$C_4(r)$, we compute 
the four correlation functions of the Ising pairs numerically. The 
%
%
%
%
%
%
%
%
%
\begin{widetext} 

\begin{table}[h]
\caption{Asymptotics of the uniform and staggered spin correlations when 
$\Delta \!>\!0$.}
\begin{tabular}{c|lll|lll}
\hline\hline 
& $\qquad{\cal S}_{I}^{x}(r)\quad$ 
& $\qquad{\cal S}_{I}^{y}(r)\quad$ 
& $\qquad\Delta{\cal S}_{I}^{z}(r)\quad$ 
& $\qquad{\cal S}_{N}^{x}(r)\quad$ 
& $\quad{\cal S}_{N}^{y}(r)\quad$ 
& $\quad{\cal S}_{N}^{z}(r)\quad$ \\ \hline
$h=0$ & $-r^{-1} e^{-(m_2+m_3)r}$ & $-r^{-1}e^{-(m_1+m_3)r}$ 
& $-r^{-1}e^{-(m_1+m_2)r}$ 
& \;\;\,$r^{-1/2}e^{-m_1r}$ & \,$r^{-1/2}e^{-m_2r}$ 
& \;\, $r^{-1/2}e^{-m_3r}$ \\ 
$0\!<\!h\!<\!h_c$ & $-r^{-1} e^{-(m_{\text{eff}}+m_3)r}$ 
& \;\, $r^{-2} e^{-(m_{\text{eff}}+m_3)r}$ 
& \;\, $r^{-2} e^{-2m_{\text{eff}}r}$ 
& $-r^{-3/2} e^{-m_{\text{eff}}r}$ & \,$r^{-1/2} e^{-m_{\text{eff}}r}$ 
& \;\, $r^{-1/2} e^{-m_3r}$ \\ 
$h=h_c$ & $-r^{-3/2} e^{-m_3r}$ & \;\, $r^{-3/2}e^{-m_3r}$ 
& \;\, $r^{-2}$ 
& $-r^{-9/4}$ & \,$r^{-1/4}$ & \;\, $r^{-3/4} e^{m_3r}$ \\ 
$h_c\!<\!h\!<\!\tilde{h}_c^*$ & $-r^{-2} e^{-(m_{\text{eff}}+m_3)r}$ 
& \;\, $r^{-1} e^{-(m_{\text{eff}}+m_3)r}$ 
& \;\, $r^{-2} e^{-2m_{\text{eff}}r}$ 
& $-r^{-3} e^{-2m_{\text{eff}}r}$ & \,$1$ 
& \;\, $r^{-1} e^{-(m_{\text{eff}} + m_3) r}$ \\ 
$h>\tilde{h}_c^*$ & 
$\begin{array}{l} \!\! -r^{-2} e^{-(m_{\text{eff}}+m_3)r} \\ 
\;\times \cos(\tilde{k}_c^*r \!+\! \delta_I^x) \end{array}$ & 
\!\!\!$\begin{array}{l} \,-r^{-1} e^{-(m_{\text{eff}}+m_3)r} \\ 
\;\;\, \times \cos(\tilde{k}_c^*r \!+\! \delta_I^y) \end{array}$ & 
\!\!\!$\begin{array}{l} \,-r^{-2} e^{-2m_{\text{eff}}r} \\ 
\;\;\, \times [1\!-\! \eta \cos(2\tilde{k}_c^*r \!-\! \delta_I^z)] \end{array}$ 
& \;\, $r^{-1} e^{-2m_{\text{eff}}r}$ & \,$1$ 
& $\begin{array}{l} \!\! -r^{-1} e^{-(m_{\text{eff}}+m_3)r} \\ 
\; \times \cos(\tilde{k}_c^*r \!+\! \delta_{\mu\mu}) \end{array}$ \\ 
\hline\hline
\end{tabular} 
\label{tab:SINcf} 
\end{table} 
\end{widetext} 
results are shown in Fig.~\ref{fig:Iscf}. We find that below the Ising 
transition ${\cal C}_{\mu\mu}^{ab}(r)$, as in the $h=0$ case, is asymptotically 
a constant, while above the transition ${\cal C}_{\mu\sigma}^{ab}(r)$ becomes 
a constant. The lines decrease faster (in fact twice) in 
${\cal C}_{\sigma\sigma}^{ab}(r)$  
before the Ising transition and in 
${\cal C}_{\sigma\mu}^{ab}(r)$ after the transition. Furthermore, even in the 
incommensurate phase, functions ${\cal C}_{\sigma\mu}^{ab}(r)$ and 
${\cal C}_{\mu\sigma}^{ab}(r)$ are not modulated by the wavevector 
$\tilde{k}_c^*$. 
We also recognize that with respect to the effective Ising theory, the 
combined operators 
$\mu_1\sigma_2$ and $\mu_1\mu_2$ take effectively the roles of the Ising order 
and disorder fields across the transition, since their correlations have the 
standard power and exponential forms.   

Finally, by Eqs.~(\ref{SI}) and (\ref{SN}), we obtain the results of spin 
correlations in various phases. In the axially symmetric case ($\Delta\!=\!0$), 
when the field exceeds the C-IC transition point, we find the transverse 
uniform spin correlations: ${\cal S}_I^{x,y}(r) 
\!\propto\! r^{-3/2} e^{-m_3r} \sin[k_cr - \delta(k_c)]$ with the phase shift 
$\delta(k_c) \!\sim\! \frac{k_c}{m}$ and the incommensurate wavevector $k_c 
\!\approx\! (\frac{2m}{\lambda})^{1/2} (h-m)^{1/2}$, and the longitudinal 
uniform spin correlations: $\Delta {\cal S}_I^{z}(r) \!=\! {\cal S}_I^{z}(r) 
- M^2 \!\propto\! -r^{-2} [1- \cos(2k_cr)]$ with $M \!=\! \frac{k_c}{\pi}$ 
the magnetization; and the transverse staggered correlations: 
${\cal S}_N^{x,y}(r) \!\propto\! r^{-1/2}$, and the longitudinal staggered 
correlations: ${\cal S}_N^{z}(r) \!\propto\! r^{-1} e^{-m_3 r} \cos(k_cr)$. 
These results are in agreement with those from other theories \cite{fur99} just 
above the transition point. In anisotropic case, the asymptotics for 
the spin correlations are summarized in Table~\ref{tab:SINcf}. Here, the phase 
shifts in the incommensurate phase, restricted within $(0,\pi)$, are in 
general field and mass dependent; $\eta (< \!1)$ is also a non-universal 
constant. As argued by Affleck \cite{aff90}, we verify that one of the 
transverse staggered spin correlations (${\cal S}_N^y$ in our case) behaves 
simply like the correlation of an Ising variable across the transition, and 
the fluctuation of the longitudinal uniform correlation obeys the power-law 
$\Delta {\cal S}_I^z \!\propto\! r^{-2}$ at the criticality. 

To summarize, we have studied the Haldane spin-1 system with generic 
anisotropies under the effect of external magnetic field. We showed explicitly 
the Ising criticality in both the spectrum and correlation functions. The 
latter was calculated based on the establishment of the block-Toeplitz 
determinant representation for the correlations of the Ising pair-fields. 
A true long-range order is formed above the transition. We also found the 
incommensurability in high field is a rather robust 
property, which appears in both iso- and anisotropic systems, and the mass 
(not the gap) is field independent in the incommensurate phase. Finally, we 
emphasize the fact that the C-IC transition (point $\tilde{h}_c^*$) in the 
massive phase is thermodynamically unidentifiable, in other words, there is 
no anomaly associated with this transition in either magnetization or specific 
heat measurement. The discussions in this respect, as well as the detailed 
formalism of the present work, will be published elsewhere. 

I am very grateful to F.\ Essler and A.\ Nersesyan for raising my interest in 
the subject and critical readings of the manuscript. The author also wishes 
to thank P.\ Fulde, A.M.\ Tsvelik, R.A.\ Klemm, and Q.\ Gu for helpful 
discussions. The work was supported under the visitors program of 
MPI-PKS.



\begin{thebibliography}{99}

\bibitem{hald83} F.D.M.\ Haldane, Phys.\ Lett.\ {\bf 93A}, 464 (1983); 
Phys.\ Rev.\ Lett.\ {\bf 50}, 1153 (1983); for a review, see I.\ Affleck, 
J.\ Phys.: Condens.\ Matter {\bf 1}, 3047 (1989). 
\bibitem{aff90} I.\ Affleck, Phys.\ Rev.\ B {\bf 41}, 6697 (1990); 
Phys.\ Rev.\ B {\bf 43}, 3215 (1991).
\bibitem{tsv90} A.M.\ Tsvelik, Phys.\ Rev.\ B {\bf 42}, 10499 (1990). 
\bibitem{sak91} T.\ Sakai and M.\ Takahashi, Phys.\ Rev.\ B {\bf 43}, 13383 
(1991). 
\bibitem{sach94} S.\ Sachdev, T.\ Senthil, and R.\ Shankar, 
Phys.\ Rev.\ B {\bf 50}, 258 (1994). 
\bibitem{fur99} A.\ Furusaki and S.-C.\ Zhang, Phys.\ Rev.\ B {\bf 60}, 
1175 (1999); 
R.M.\ Konik and P.\ Fendley, Phys.\ Rev.\ B {\bf 66}, 
144416 (2002). 
\bibitem{ren87} J.P.\ Renard {\it et al}., Europhys.\ Lett.\ {\bf 3}, 945 (1987).
\bibitem{aji89} Y.\ Ajiro, T.\ Goto, H.\ Kikuchi, T.\ Sakakibara, and T.\ Inami, 
Phys.\ Rev.\ Lett.\ {\bf 63}, 1424 (1989). 
\bibitem{hon01} Z.\ Honda, K.\ Katsumata, Y.\ Nishiyama, and I.\ Harada, 
Phys.\ Rev.\ B {\bf 63}, 64420 (2001). 
\bibitem{hon98} Z.\ Honda, H.\ Asakawa, and K.\ Katsumata, 
Phys.\ Rev.\ Lett.\ {\bf 81}, 2566 (1998). 
\bibitem{chi91} M.\ Chiba, Y.\ Ajiro, H.\ Kikuchi, T.\ Kubo, and 
T.\ Morimoto,  Phys.\ Rev.\ B {\bf 44}, 2838 (1991). 
\bibitem{chen01} Y.\ Chen {\it et al}., Phys.\ Rev.\ Lett.\ {\bf 86}, 1618 (2001);  
A.\ Zheludev {\it et al}., {\it ibid}.\ {\bf 88}, 077206 (2002); 
A.\ Zheludev {\it et al}., preprint (cond-mat/0301424).
\bibitem{dzh78} G.I.\ Dzhaparidze and A.A.\ Nersesyan, 
Pis'ma Zh.\ Eksp.\ Teor.\ Fiz.\ {\bf 27}, 356 (1978) [JETP 
Lett.\ {\bf 27}, 334 (1978)]; V.L.\ Pokrovsky and A.L.\ Talapov, 
Phys.\ Rev.\ Lett.\ {\bf 42}, 65 (1979); H.J.\ Schulz, Phys.\ Rev.\ B 
{\bf 22}, 5274 (1980). 

\end{thebibliography}
\end{document}